# Silicon Integrated Photonic Waveguide Polarizers with 2D MoS₂ Films

*Junkai Hu, Member, IEEE, Jiayang Wu, Senior Member, IEEE, Irfan H. Abidi,*
*Di Jin, Member, IEEE, Yuning Zhang, Jianfeng Mao, Anchal Pandey,*
*Yijun Wang, Sumeet Walia, and David J. Moss, Life Fellow, IEEE*

(Invited paper)

*Abstract*—Polarization control is of fundamental importance for modern optical systems, and optical polarizers serve as critical components for enabling this functionality. Here, we experimentally demonstrate optical polarizers by integrating 2D molybdenum disulfide (MoS$_2$) films onto silicon photonic waveguides. High-quality monolayer MoS$_2$ films with highly anisotropic light absorption are synthesized via a low-pressure chemical vapor deposition (LPCVD) method and subsequently transferred onto silicon-on-insulator (SOI) nanowire waveguides to fabricate integrated optical polarizers. Detailed measurements are carried out for the fabricated devices with various MoS$_2$ film coating lengths and silicon waveguide geometry. The results show that a maximum polarization-dependent loss of ~21 dB is achieved, together with a high figure of merit of ~4.2. In addition, the hybrid waveguide polarizers exhibit broad operation bandwidth exceeding ~100 nm and excellent power durability. These results highlight the strong potential for on-chip integration of 2D MoS$_2$ films to implement high-performance polarization selective devices.

## I. INTRODUCTION

The control of light polarization plays a crucial role in modern optical systems and underpins a wide range of advanced optical technologies [1, 2]. Optical polarizers, which allow transmission of light in a specific polarization while suppressing the orthogonal polarization component, serve as key elements for polarization control in optical systems [3]. To date, a variety of optical polarizers have been developed based on refractive prisms [4, 5], birefringent crystals [6, 7], fiber components [8, 9], and integrated photonic devices [10, 11]. However, these polarizers based on bulk materials often struggle to achieve efficient polarization selection over wide wavelength ranges. This limitation is especially significant given the increasing demand for broadband optical polarizers driven by rapid progress in photonic technologies and systems [12, 13].

Recently, due to strong anisotropy in light absorption and broadband response, two-dimensional (2D) materials with atomic-scale film thicknesses have been incorporated onto bulk optical waveguides to realize high-performance optical polarizers [14], as demonstrated by those incorporating 2D materials such as graphene [14-16], graphene oxide (GO) [17-19], and transition metal dichalcogenides (TMDCs) [20-22]. As a significant subgroup in the 2D material family, TMDCs exhibit a direct bandgap in their monolayer form that transitions to an indirect bandgap in few-layer or bulk forms [23, 24]. This distinctive property facilitates their widespread applications for next-generation atomically thin devices such as transistors [25, 26], photodetectors [27, 28], and electrocatalysts [29, 30]. More recently, the strong anisotropic light absorption of 2D TMDC films across broad wavelength ranges has also been explored, and their integration onto polymer and Neodymium-doped Yttrium Aluminum Garnet (Nd:YAG) waveguides for realizing optical polarizers has been successfully demonstrated [21, 22].

In this work, we demonstrate the integration of 2D molybdenum disulfide (MoS$_2$) – a representative TMDC material – onto the widely used silicon photonic platform to realize high-performance optical polarizers. High-quality monolayer MoS$_2$ films with strong anisotropy in light absorption are synthesized via a low-pressure chemical vapor deposition (LPCVD) method, and subsequently transferred onto silicon-on-insulator (SOI) nanowire waveguides using polymer-assisted transfer process. We perform detailed measurements for the fabricated devices with different MoS$_2$ film coating lengths and silicon waveguide geometry, achieving a maximum polarization-dependent loss of ~21 dB and a high figure of merit of ~4.2. In addition, the hybrid waveguide polarizers exhibit broad operation bandwidth over ~100 nm and excellent power durability. Finally, we compare the

Jiayang Wu and David J. Moss are with the Optical Sciences Center, Swinburne University of Technology, Melbourne, 3122, Australia, and with the ARC Centre of Excellence in Optical Microcombs for Breakthrough Science (COMBS), Melbourne, Victoria 3000, Australia (e-mail: jiayangwu@swin.edu.au; dmoss@swin.edu.au).
Irfan H. Abidi, Jianfeng Mao, Anchal Pandey, and Sumeet Walia are with the School of Engineering, RMIT University, 124 La Trobe Street, Melbourne 3000, Australia, and with the ARC Centre of Excellence in Optical Microcombs for Breakthrough Science (COMBS), (e-mail: Irfan.haider.abidi@rmit.edu.au; jianfeng.mao@rmit.edu.au; S4077929@student.rmit.edu.au; sumeet.walia@rmit.edu.au).
Yuning Zhang is with the School of Physics, Peking University, Beijing, 100871, China (e-mail: yuningzhang0731@pku.edu.cn)
Yijun Wang is with the School of Automation, Central South University, Changsha, 410083, China (e-mail: xxywyj@csu.edu.cn).

*Correspondence: Jiayang Wu; Irfan H. Abidi; Sumeet Walia; David J. Moss.*
Junkai Hu and Di Jin are with the Optical Sciences Center, Swinburne University of Technology, Melbourne, 3122, Australia, with the School of Automation, Central South University, Changsha, 410083, China, and also with the ARC Centre of Excellence in Optical Microcombs for Breakthrough Science (COMBS), Melbourne, Victoria 3000, Australia (e-mail: junkaihu@swin.edu.au; dijin@swin.edu.au).

performance of our device with state-of-the-art waveguide polarizers incorporating different 2D materials and find that it achieves the highest figure of merit (FOM) among all those based on the silicon photonic platform. These results reveal the strong potential of 2D TMDC films for implementing high-performance integrated polarization selective devices.

## II. DEVICE DESIGN

As an important member of the TMDCs family that has been widely studied, $MoS_2$ features a hexagonal sheet of molybdenum (Mo) atoms sandwiched between two hexagonal sheets of sulfur (S) atoms [31, 32]. **Fig. 1(a)** illustrates the atomic structure of monolayer $MoS_2$, where Mo and S atoms are connected by strong covalent bonds. Multi-layered $MoS_2$ consists of vertically stacked layers that are weakly bonded through van der Waals interactions [26]. Monolayer $MoS_2$ has shown great potential as a semiconducting material with a direct bandgap of ~1.8 – 1.9 eV [33]. This is larger than the energy of two photons at 1550 nm (*i.e.*, ~1.6 eV), which allows for relatively low linear light absorption as well as two-photon absorption at near infrared wavelengths. The quality of 2D $MoS_2$ crystals plays a crucial role in determining the material properties such as refractive index, light absorption, and optical bandgap, which correlates to the intrinsic structural defects induced by sulfur vacancies [34, 35]. In this work, we choose $MoS_2$ to implement optical polarizers due to several compelling advantages. First, it exhibits strong anisotropic light absorption over a very broad spectral bandwidth [36, 37]. Second, it possesses relatively low linear absorption in the infrared regime, with an extinction coefficient nearly 1 order of magnitude lower than that of graphene [38, 39], making it particularly suitable for infrared photonic applications. Finally, we develop a simple and one-step chemical vapor deposition (CVD) method to fabricate $MoS_2$ films with precise control over structural defects [34], which allows us to tailor the intrinsic optical properties of $MoS_2$ for specific optical applications.

**Fig. 1(b)** shows the schematic of an integrated waveguide polarizer consisting of a silicon nanowire waveguide coated with a monolayer $MoS_2$ film. Similar to other 2D materials such as graphene and graphene oxide (GO) [14, 15, 17, 19], 2D $MoS_2$ films exhibit strong anisotropic optical absorption [40, 41], with significantly higher absorption for light propagating in the in-plane direction compared to the out-of-plane direction. For the hybrid waveguide in **Fig. 1(b)**, these directions correspond to TE- and TM-polarized incident light, respectively. As a result, the hybrid waveguide can effectively operate as a TM-pass waveguide polarizer. It is worth noting that 2D $MoS_2$ exhibits a broad spectral range of material anisotropy spanning from visible to infrared wavelengths [36]. This wide bandwidth offers a significant advantage for $MoS_2$-coated integrated waveguide polarizers, which is difficult to achieve for conventional bulk silicon photonic polarizers [1, 14, 42].

**Fig. 1(c)** shows a schematic of the cross section of the hybrid waveguide in **Fig. 1(b)**. The corresponding transverse electric (TE) and transverse magnetic (TM) mode profiles at 1550 nm are provided in **Fig. 1(d)**, which were simulated using commercial mode-solving software (COMSOL Multiphysics). In our simulation, the thickness of the monolayer $MoS_2$ film was ~0.7 nm. The refractive index (*n*) and extinction coefficient (*k*) of $MoS_2$ for TE polarization were $n_{TE}$ = ~3.8 and $k_{TE}$ = ~0.107, respectively. For TM polarization, the corresponding values were $n_{TM}$ = ~3.2 and $k_{TM}$ = ~0.027. These values were obtained from our measurements in the following sections. The simulated TE- and TM polarized effective indices for the hybrid waveguide were ~2.081 + 3.313 × $10^{-4}$i and ~1.551 + 7.132 × $10^{-5}$i, respectively. The large difference in the imaginary part highlights the polarization selectivity for the hybrid waveguide, which originates from the significant disparity between $k_{TE}$ and $k_{TM}$ of the 2D $MoS_2$ film. It should be noted that, due to the polymer-assisted transfer method used in our fabrication process (which will be discussed in **Section III**), the monolayer $MoS_2$ film does not conformally coat on the sidewalls of the silicon nanowire waveguide, resulting in air gaps between the waveguide sidewalls and the $MoS_2$ film. Nevertheless, this has minimal impact on the polarization selectivity of the hybrid waveguides, as it mainly depends on the interaction between the evanescent field and the $MoS_2$ film on the waveguide top surface.

## III. DEVICE FABRICATION AND MATERIAL CHARACTERIZATION

Based on the device design in Section **II**, we fabricated silicon photonic waveguide polarizers coated with monolayer $MoS_2$ films in this section. In addition, we employed a range of material characterization methods to assess the quality of the $MoS_2$ films on the photonic integrated chips.

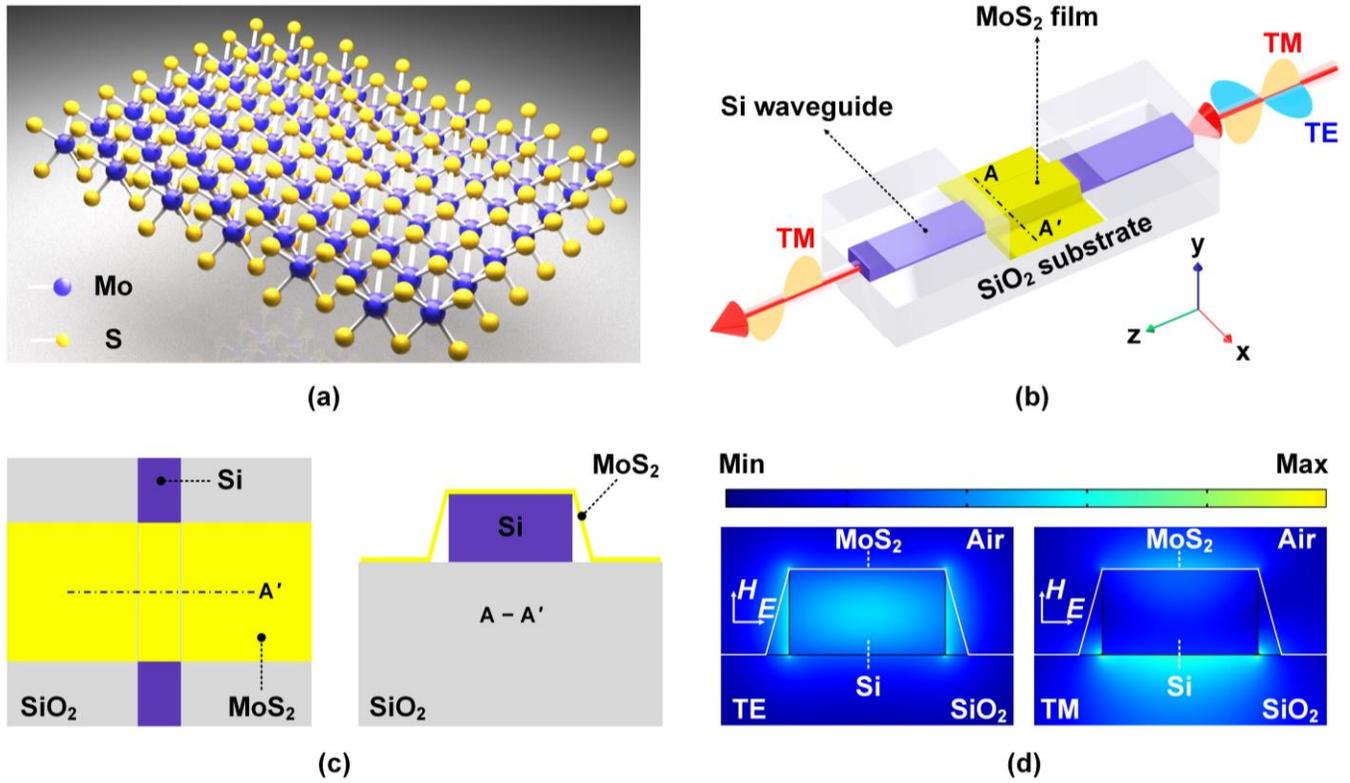

Fig. 1. (**a**) Schematic atomic structure of monolayer molybdenum disulfide (MoS$_2$). (**b**) Schematic illustration of a silicon (Si) nanowire waveguide integrated with a monolayer MoS$_2$ film as a waveguide polarizer. (**c**) Schematic illustrations of top view and cross section of the hybrid waveguide in (**b**). (**d**) TE and TM mode profiles for the hybrid waveguide in (**c**).

We first fabricated uncoated silicon nanowire waveguides using CMOS-compatible fabrication technologies. The nanowire waveguides were fabricated on silicon-on-insulator (SOI) wafer with a 220-nm-thick top silicon layer and a 2-μm-thick silica layer. The waveguide patterns were defined using 248-nm deep ultraviolet photolithography, followed by waveguide formation through an inductively coupled plasma etching process. After this, a 1.5-μm-thick silica layer serving as an upper cladding layer was deposited on the SOI chip via plasma enhanced chemical vapor deposition (PECVD). Finally, windows of different lengths were opened on the silica upper cladding through the processes of photolithography and reactive ion etching (RIE) to enable the coating of MoS$_2$ films onto the silicon waveguides. All the nanowire waveguides we fabricated had the same length of ~3.0 mm, and the lengths of the opened windows (*i.e.*, the MoS$_2$ film coating lengths) ranged between ~0.1 mm and ~2.2 mm.

After fabricating the uncoated silicon waveguides, we used an LPCVD method that we developed in Ref. [34] to synthesize high-quality MoS$_2$ films. Our method allows for precise control of the intrinsic atomic defects induced by sulfur vacancies, and supports the direct growth of high-quality and large-area 2D MoS$_2$ films with precise thickness control. **Fig. 2(a)** shows a schematic of our CVD process flow using a two-temperature-zone tube furnace to synthesize MoS$_2$ monolayers. First, the Mo precursor was drop-casted onto an ultrasonically cleaned substrate in Zone 2 of the furnace, which was maintained at 750 °C during the CVD process. After this, S powder was vaporized at 180 °C in Zone 1, and the S vapors were carried downstream by a 70 sccm flow of argon as carrier gas into Zone 2. Finally, a surface reaction between the Mo and S species occurred, resulting in the synthesis of MoS$_2$ single crystals. During the growth, the process pressure was maintained at ~1 torr to maintain a low-pressure CVD conditions. **Fig. 2(b)** shows a microscopic image of a monolayer MoS$_2$ film grown on a sapphire substrate, which exhibits a high film uniformity. **Fig. 2(c)** shows an atomic force microscopy (AFM) image of a representative MoS$_2$ crystal synthesized by using our CVD method, which shows an average thickness of ~0.7 nm.

Following the CVD synthesis of a monolayer MoS$_2$ film, it was transferred onto the SOI chip with uncoated silicon waveguides using a polymer-assisted transfer process [43]. The as-grown MoS$_2$ film was initially spin-coated with a polystyrene (PS) support layer. Subsequently, the PS/MoS$_2$ stack was exfoliated from the growth substrate using a water-droplet-assisted delamination process, driven by surface energy differences between the interfaces [44]. Finally, the resulting stack was then stamped onto the SOI chip via van der Waals interactions, and the PS layer was removed by dissolving it in toluene. As shown in **Fig. 2(d)**, the transferred monolayer MoS$_2$ film exhibits high optical transmittance and uniform coverage across the SOI chip surface, confirming the effectiveness and quality of the transfer process.

**Fig. 3(a-i)** and **(a-ii)** show the Raman spectra of the same SOI chip before and after the transfer of monolayer MoS$_2$ film, which were measured by using a ~514-nm excitation laser. After MoS$_2$ integration, the Raman spectrum exhibits two prominent peaks emerging at ~384 cm$^{-1}$ and ~404 cm$^{-1}$, corresponding to the in-plane ($E^1_{2g}$) TE and out-of-plane ($A_{1g}$) TM vibrational modes of monolayer MoS$_2$, respectively.

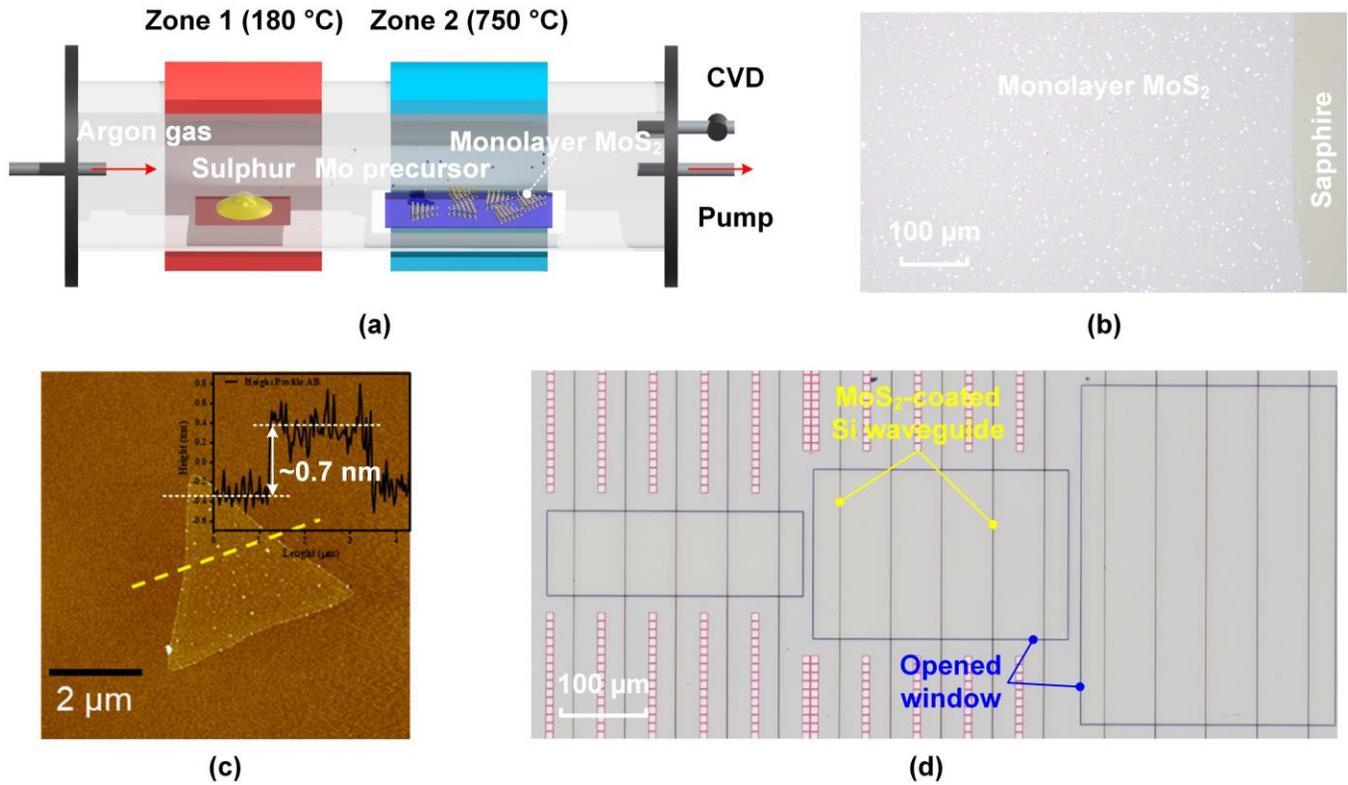

Fig. 2. (**a**) Schematic illustration showing the chemical vapor deposition (CVD) process flow we employed for synthesizing MoS$_2$ monolayers. (**b**) Microscopic image of a monolayer MoS$_2$ film coated on a sapphire substrate. (**c**) Atomic force microscopy (AFM) image of a MoS$_2$ crystal synthesized by using the CVD method in (**a**). (**d**) Microscopic image of a MoS$_2$-coated silicon-on-insulator (SOI) chip with opened windows.

Moreover, the observed frequency difference of ~20 cm$^{-1}$ is characteristic of monolayer MoS$_2$. An additional peak observed at ~517 cm$^{-1}$ originates from the underlying silicon substrate. These spectral features are consistent with those reported previously for CVD-grown monolayer MoS$_2$ [34, 45, 46], confirming successful and high-quality integration of 2D MoS$_2$ onto the SOI chip.

**Fig. 3(b)** shows the spectra for linear optical absorption and transmittance of the synthesized MoS$_2$ film, which were characterized via ultraviolet–visible (UV–vis) spectrometry. It can be seen that the MoS$_2$ exhibited strong light absorption in the visible and infrared wavelength regions. The linear absorption spectrum exhibited a sharp increase followed by a rapid decline within the range of ~400 – 600 nm. Two strong absorption peaks were observed at ~605 cm$^{-1}$ and ~651 cm$^{-1}$, and the peak associated with van Hove singularities [47] of monolayer MoS$_2$ was also observed at ~430 cm$^{-1}$. Before a gradual decrease at wavelengths >900 nm, the linear absorption spectrum exhibited a sudden jump at ~870 nm. The transmittance of the sample had a transmittance >60% at wavelengths between ~400 nm and ~1800 nm. These results show agreement with those reported in previous studies [48-50] and further validate the high quality of our synthesized 2D MoS$_2$ films.

**Fig. 3(c)** shows the X-ray diffraction (XRD) spectrum for the synthesized MoS$_2$ film. The diffraction rings can be indexed to the (100), (103), (105), and (110) reflections of hexagonal MoS$_2$ crystal structure, which shows an agreement with the measured XRD spectra of MoS$_2$ in Refs. [51, 52]. **Fig. 3(d)** shows the X-ray photoelectron spectroscopy (XPS) analysis of the synthesized MoS$_2$ film. In **Fig. 3(d-i)**, prominent peaks were observed at ~226.9 eV, ~229.6 eV, and ~232.7 eV, which correspond to S 2$s$, Mo 3$d$ doublet of Mo$^{4+}$ 3$d_{5/2}$ and Mo$^{4+}$ 3$d_{3/2}$. In addition, weak peaks at ~232.3 eV and ~235.4 eV were attributed to the $3d_{5/2}$ and $3d_{3/2}$ components of Mo$^{6+}$, indicating the presence of MoO$_x$ species, likely originating from partial surface oxidation or physiosorbed oxygen during ambient exposure [35]. **Fig. 3(d-ii)** displays the *S 2p* region, where two peaks appeared at ~162.6 eV and ~163.8 eV and corresponded to the S 2$p_{3/2}$ and S 2$p_{1/2}$, respectively, further supporting the formation of MoS$_2$ with the expected stoichiometry. These experimental results are consistent with previously reported XPS signatures of CVD-grown monolayer MoS$_2$ in Refs. [46, 53-55].

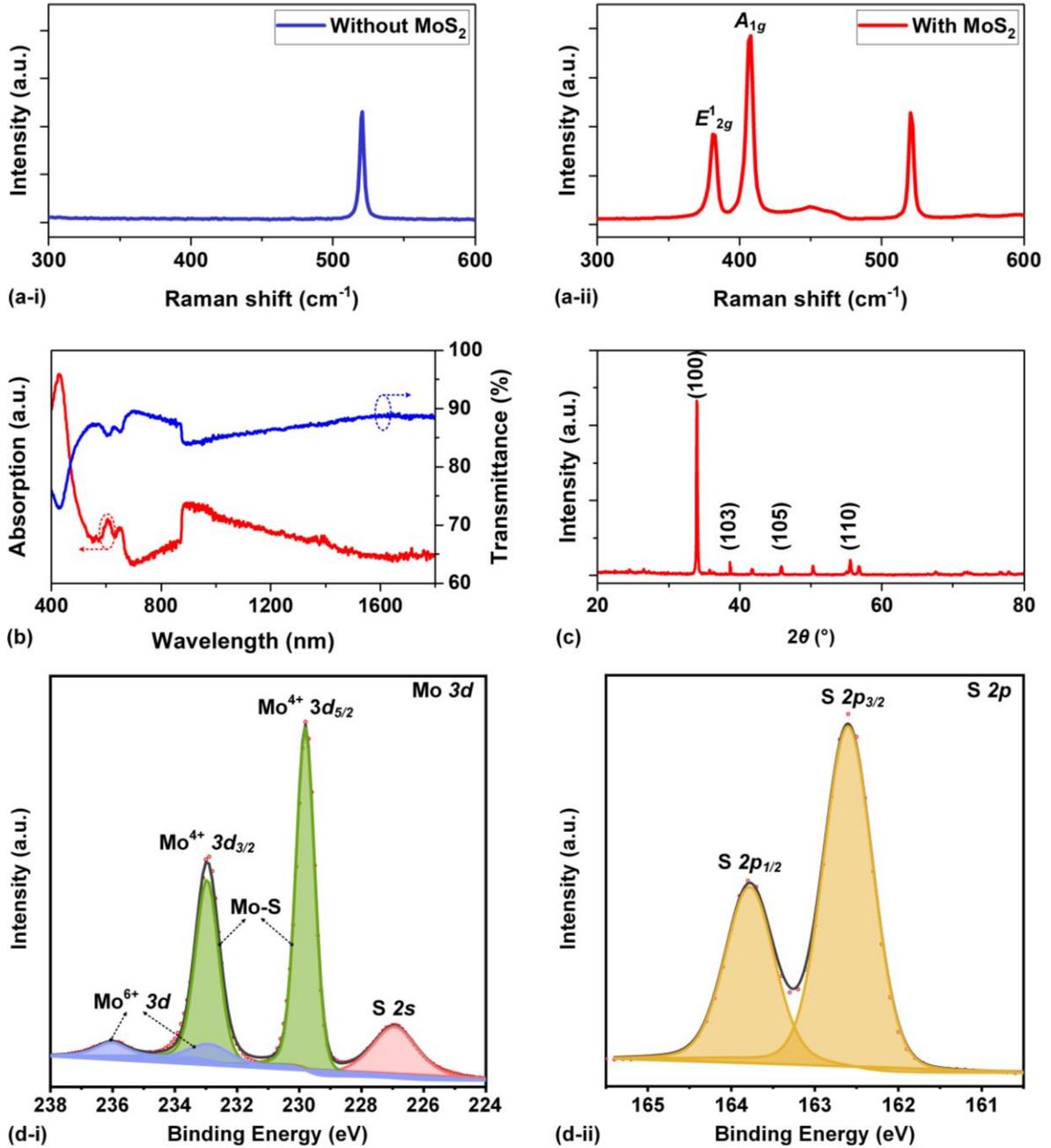

Fig. 3. Characterization of our synthesized monolayer $MoS_2$ films. (**a**) Measured Raman spectra of an SOI chip (**i**) before and (**ii**) after coating a monolayer $MoS_2$ film. (**b**) Ultraviolet–visible (UV–vis) absorption and transmittance spectra. (c) X-ray diffraction (XRD) spectrum. (**d**) X-ray photoelectron spectroscopy (XPS) spectra, where (**i**) and (**ii**) show characteristic peaks of Mo $3d$ and S $2p$, respectively.

## IV. POLARIZATION-DEPENDENT LOSS MEASUREMENTS

In this section, we measured the polarization-dependent loss (*PDL*) of the fabricated $MoS_2$-Si hybrid waveguides in **Section III** for input continuous-wave (CW) light with different polarization states. We performed measurements for devices with various silicon waveguide widths (*W*) and with monolayer $MoS_2$ films of different coating lengths ($L_c$). In our measurements, lensed fibers were employed to butt couple a CW light at ~1550 nm into and out of the fabricated devices with inverse-taper couplers at both ends. The fiber-to-chip coupling loss was ~5 dB / facet.

**Fig. 4(a-i)** plot the measured TE- and TM-polarized insertion loss (*IL*) versus $MoS_2$ film coating length $L_c$ for the hybrid waveguides with monolayer $MoS_2$ films. For comparison, all the devices had the same $W$ = ~400 nm. During our measurements in **Fig. 4**, the input CW power and wavelength were kept the same as $P_{in}$ = ~0 dBm and $\lambda$ = ~1550 nm, respectively. Unless otherwise specified, the values of $P_{in}$ and *IL* in our following discussions refer to those after excluding the fiber-to-chip coupling loss.

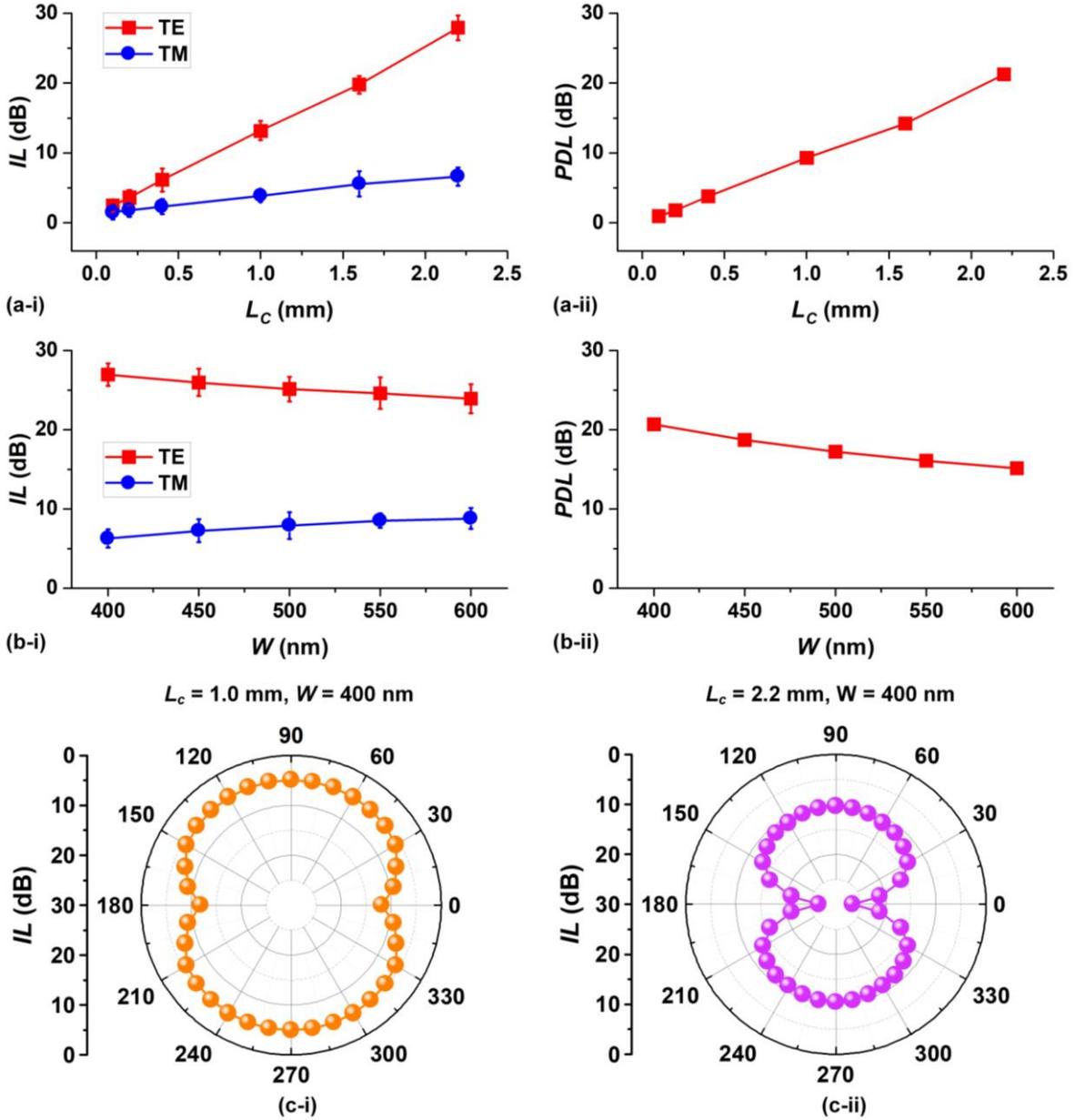

Fig. 4. (**a**) Measured *IL* versus MoS$_2$ film coating length ($L_c$) for the MoS$_2$-Si hybrid waveguides. (**b**) Measured insertion loss (*IL*) versus the waveguide width (*W*) for the MoS$_2$-Si hybrid waveguides. In (**a**) and (**b**), (**i**) shows the measured results for TE & TM polarizations, and (**ii**) shows the polarization dependent loss (*PDL*) calculated from (**i**). (**c**) Polar diagrams for the measured *IL* of devices with (**i**) $L_c$ = 1.0 mm and (**ii**) $L_c$ = 2.2 mm. The polar angle represents the angle between the input polarization plane and the substrate. In (**a-c**), the input continuous-wave (CW) power and wavelength were $P_{in}$ = ~0 dBm and $\lambda$ = ~1550 nm, respectively. In (**c**), $W$ = ~400 nm.

In **Fig. 4(a-i)**, the data points depict the average of measurements on three duplicate devices, and the error bars reflect the variations among different devices. As can be seen, the *IL* increases with $L_c$ for both TE and TM polarizations, with the former exhibiting a faster rate of increase than the latter. This reflects a higher propagation loss for TE polarization, which is associated with a larger imaginary part of its effective index, as simulated in **Fig. 1(d)**.

In **Fig. 4(a-ii)**, we further calculated the *PDL* (dB) by subtracting the TM-polarized *IL* from the TE-polarized *IL* in **Fig. 4(a-i)**. For the device with $L_c$ = 2.2 mm, a maximum *PDL* value of ~21 dB was achieved. In contrast, the uncoated Si waveguide did not show any significant polarization-dependent *IL*, with a *PDL* below 0.5 dB. The huge difference in the *PDL* values highlights the polarization selectivity introduced by integrating a 2D MoS$_2$ film onto the silicon photonic waveguide. In **Fig. 4(a-i)**, the *PDL* increases with $L_c$, this further confirms that the exceptional polarization selectivity arises from the 2D MoS$_2$ film and suggests that improved *PDL* can be achieved by increasing the MoS$_2$ film coating length.

**Fig. 4(b-i)** shows the measured TE- and TM-polarized *IL* versus waveguide width *W* for the hybrid waveguides with the same $L_c$ = ~2.2 mm. In **Fig. 4(b-i)**, the *IL* increases with *W* for TM polarization but decreases for TE polarization. This is mainly resulting from changes in the mode overlap of TE and TM modes induced by variations in the silicon waveguide

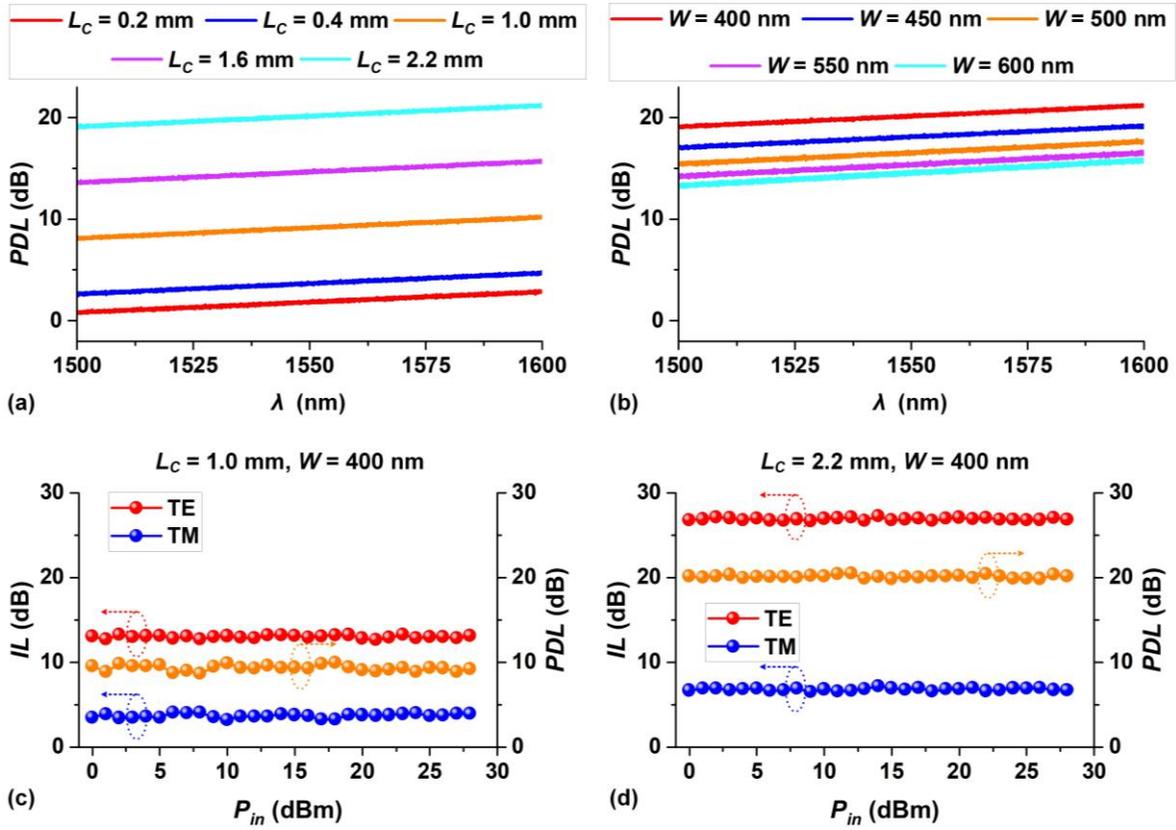

Fig. 5. (**a**) Measured *PDL* versus input CW wavelength $\lambda$ for the MoS$_2$-Si hybrid waveguides with different $L_c$ but the same $W$ of ~400 nm. (**b**) Measured *PDL* versus $\lambda$ for the MoS$_2$-Si hybrid waveguides with different $W$ but the same $L_c$ of ~2.2 mm. (**c**) – (**d**) Measured TE- and TM-polarized *IL* and calculated *PDL* versus input power $P_{in}$ for the MoS$_2$-Si hybrid waveguides with $L_c$ = ~1.0 mm and ~2.2 mm, respectively. In (**a**) and (**b**), $P_{in}$ = ~0 dBm. In (**c**) and (**d**), the input CW wavelength and the waveguide width were $\lambda$ = ~1550 nm and $W$ = ~400 nm, respectively.

width. **Fig. 4(b-ii)** shows the corresponding *PDL* calculated from the measured *IL* in **Fig. 4(b-i)**. The device with $W$ = ~400 nm achieved a maximum *PDL* of ~21 dB, and the *PDL* decreased as $W$ increased, reaching ~15 dB at $W$ = ~600 nm. **Fig. 4(c)** shows the polar diagrams for the measured *IL* of devices with the same $W$ = ~400 nm but different $L_c$ = ~1.0 mm and ~2.2 mm. In the polar diagrams, the variations in the *IL* values across different polarization angles further confirm the polarization selectivity of the MoS$_2$-Si hybrid waveguides.

In addition to measuring *PDL* at constant wavelength ($\lambda$ = ~1550 nm) and power ($P_{in}$ = ~0 dBm) for the input CW light in **Fig. 4**, we also investigated the dependence of the *PDL* on input light wavelength and power. **Fig. 5(a)** shows the measured *PDL* versus input CW wavelength $\lambda$ for the hybrid waveguides with various $L_c$ but the same $W$ = 400 nm. During our measurements, the input CW power was maintained at $P_{in}$ = ~0 dBm. For all the devices, there were no obvious changes in the *PDL* (< 2 dB) within the measured wavelength range of ~1500 – 1600 nm. We also notice that there was a minor increase in the *PDL* as $\lambda$ increased, which can be attributed to a slight change in MoS$_2$'s mode overlap induced by dispersion. **Fig. 5(b)** shows the measured *PDL* versus $\lambda$ for the hybrid waveguides with various $W$ but the same $L_c$ = ~2.2 mm. Similarly, there were no significant variations in the *PDL*, with only a slight increase in the *PDL* as $\lambda$ increased.

The results in **Figs. 5(a)** and (**b**) highlight the broadband operation of the MoS$_2$-Si waveguide polarizers – a feature that is often challenging to achieve for bulk silicon photonic polarizers [3, 56]. In our measurements, the wavelength tuning range was limited by the tunable CW laser employed to scan the transmission spectra. In fact, MoS$_2$ films exhibit a broad bandwidth for anisotropic light absorption, which extends well beyond that demonstrated here and can span from the visible to the infrared wavelength regions [36, 37].

**Figs. 5(c)** and (**d**) show the measured TE- and TM-polarized *IL* and calculated *PDL* versus input CW power $P_{in}$ for the hybrid waveguides with the same $W$ = ~400 nm but different $L_c$ = ~1.0 mm and ~2.2 mm, respectively. For comparison, the input CW wavelength was kept the same at $\lambda$ = ~1550 nm. In both figures, there are no notable changes in the *IL* and *PDL* within the input power range of ~0 dBm – ~28 dBm. This indicates excellent thermal stability of the 2D MoS$_2$ films and remarkable power durability of the hybrid devices. In contrast, 2D GO films coated on silicon waveguides are susceptible to photothermal reduction at $P_{in}$ > 10 dBm, as observed in our previous measurements [57]. We did not perform measurements for $P_{in}$ > ~28 dBm, as $P_{in}$ = ~28 dBm was the maximum available from our experimental setup. This value accounts for a 5-dB coupling loss subtracted from the 33-dBm maximum CW power, which was achieved after amplification by an erbium-doped fiber amplifier (EDFA).

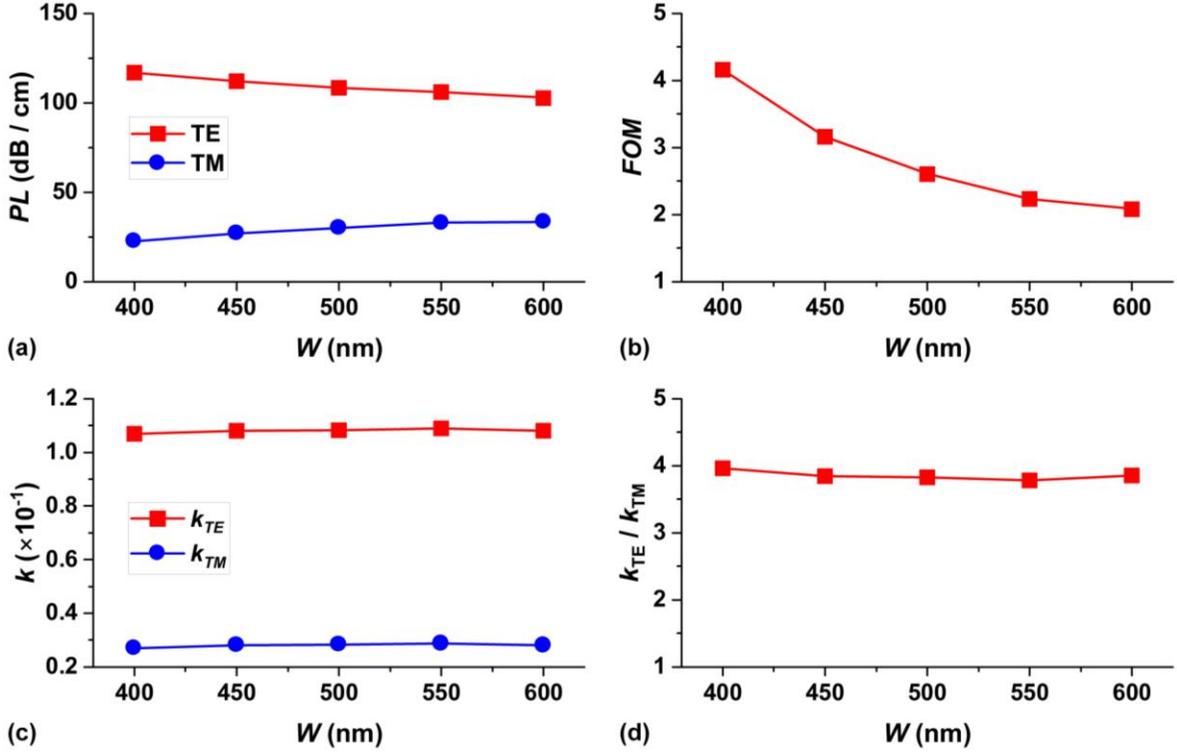

Fig. 6. (a) TE- and TM-polarized propagation loss (*PL*) versus silicon waveguide width *W* for the MoS$_2$-Si hybrid waveguides. (b) TE- and TM-polarized extinction coefficients of MoS$_2$ ($k_{TE}$, $k_{TM}$) versus *W* obtained by fitting the results in (a) with optical mode simulations. (c) Anisotropy ratio ($k_{TE}$ / $k_{TM}$) calculated from (b). (d) Calculated figures of merit (*FOM*) versus *W* for the MoS$_2$-Si hybrid waveguides.

## V. DISCUSSION

In this section, we further analyze the anisotropic absorption of 2D MoS$_2$ films by fitting the experimental results in **Section IV** with theoretical simulations. We also compare the performance of our MoS$_2$-Si waveguide polarizers with waveguide polarizers incorporating other 2D materials.

**Fig. 6(a)** shows the waveguide propagation loss (*PL*) versus *W* for both TE and TM polarizations, which was extracted from the measured *IL* in **Fig. 4(b)**. As expected, the TE-polarized *PL* is much higher than the corresponding TM-polarized *PL*. An increase in *W* leads to a smaller difference between the two, mainly caused by altered mode overlap with the MoS$_2$ films. The excess propagation loss (*EPL*) induced by the MoS$_2$ film was further calculated by excluding the *PL* for the uncoated silicon waveguide, which were ~3.4 dB/cm and ~3.0 dB/cm for TE and TM polarizations, respectively. At *W* = 400 nm, the TE-polarized *EPL* was ~117 dB/cm, significantly higher than the TM-polarized *EPL* of ~25 dB/cm, showing agreement with those reported in Ref. [58]. We also note that the value of ~117 dB/cm is more than an order of magnitude lower than the *EPL* induced by monolayer graphene coated on a silicon waveguide (*i.e.*, ~2000 dB/cm [39, 59]), yet ~5 times higher than that induced by monolayer GO (*i.e.*, ~20 dB/cm [60]).

**Fig. 6(b)** shows the extinction coefficient of MoS$_2$ ($k_{TE}$, $k_{TM}$ for TE and TM polarizations, respectively) obtained by fitting the results in **Fig. 6(a)** with optical mode simulations of the hybrid waveguides (at 1550 nm). At *W* = ~400 nm, $k_{TE}$ is ~0.107, which is about ~4 times that of $k_{TM}$. For all different *W*, the *k* values for TE polarization are significantly higher than those for TM polarization, highlighting the strong anisotropic light absorption of 2D MoS$_2$ films. For both polarizations, no significant variations in *k* were observed as *W* increased. This indicates that the changes in *PL* with *W* observed in **Fig. 6(a)** are mainly caused by variations in mode overlap, rather than differences in the MoS$_2$ film properties, highlighting the consistency of our measurements and uniformity of the coated MoS$_2$ film.

In **Fig. 6(c)**, we further plot the anisotropy ratios defined as the ratios of the corresponding *k* values for TE and TM polarizations ($k_{TE}$ / $k_{TM}$) in **Fig. 6(b)**. As can be seen, the anisotropy ratio also remained relatively consistent without any significant variations. A maximum anisotropy ratio of ~4.0 is achieved for monolayer MoS$_2$ films, which is close to ~4.5 reported for monolayer GO films [18].

To evaluate the performance of 2D-material-based optical polarizers, the figure of merit (*FOM*), defined by the following equation, is commonly used [14, 17],

$$FOM = PDL / EIL \qquad (1)$$

where *PDL* (dB) is the ratio of the maximum to minimum *IL*'s as we discussed in **Figs. 4(a-ii)** and **(b-ii)**, and *EIL* (dB) is the minimum insertion loss induced by the MoS$_2$ film over the uncoated waveguide. Note that the *EIL* accounts only for the *IL* induced by the MoS$_2$ film. In our case, it represents the excess MoS$_2$-induced *IL* for the TM polarization, as the TM mode exhibited a lower *IL*. **Fig. 6(d)** shows the calculated *FOM* versus silicon waveguide width *W*. A higher *FOM* value is achieved for a lower *W*. A maximum *FOM* of ~4.2 was achieved at *W* = ~400 nm, which decreased to ~2.0 as *W* increased to ~600 nm.

TABLE I. COMPARISON OF WAVEGUIDE OPTICAL POLARIZERS INCORPORATING 2D MATERIALS

| 2D material | Waveguide material | 2D material thickness | *WD* (µm) | *PDL* (dB) | *OBW* (µm) | *FOM* | Ref. |
|---|---|---|---|---|---|---|---|
| Graphene | Polymer | – [a] | 7.00 × 5.00 | ~19 | – [a] | ~0.7 | [16] |
| Graphene | Glass | – [a] | 11.50 × 2.60 | ~27 | ~1.23–1.61 | ~3.0 | [61] |
| Graphene | Chalcogenide | Monolayer | – [a] | ~23 | ~0.94–1.60 | ~28.8 | [14] |
| Graphene | Polymer | > or < 10 nm [b] | 10.00 × 5.00 | ~6 | – [a] | ~0.7 | [15] |
| $MoS_2$ | Nd:YAG | ~6.5 nm | – [a] | ~3 | – [a] | ~7.5 | [21] |
| $MoS_2$ | Polymer | ~2.5 nm | 8.00 × 8.00 | ~12.6 | ~0.65–0.98 | – [a] | [22] |
| GO | Polymer | ~2000 nm | 10.00 × 5.00 | ~40 | ~1.53–1.63 | ~6.2 | [19] |
| GO | Doped silica | ~2–200 nm | 3.00 × 2.00 | ~54 | ~0.63–1.60 | ~7.2 | [17] |
| GO | Silicon | ~10 nm | 0.40 × 0.22 | ~17 | ~1.50–1.60 | ~1.7 | [18] |
| rGO | Silicon | Monolayer | 0.40 × 0.22 | ~47 | ~1.50–1.60 | ~3.0 | [62] |
| $MoS_2$ | Silicon | Monolayer | 0.40 × 0.22 | ~21 | ~1.50–1.60 | ~4.2 | This work |

*WD*: waveguide dimension, *PDL*: polarization dependent loss, *OBW*: operational bandwidth, *IL*: insertion loss, *FOM*: figure of merit.
[a]There is no reported value for this parameter in the literature.
[b]The polymer waveguides with a few-layer graphene film (< 10 nm) and a thicker graphene film (> 10 nm) worked as TM- and TE-pass optical polarizers, respectively.

In **Table I**, we summarize state-of-the-art waveguide optical polarizers incorporating 2D materials and compare their performance. Here we only show the results for experimental works, and compare the key performance parameters including *PDL*, operation bandwidth (*OBW*), and *FOM*. Among the different polarizers, our work here marks the first demonstration of implementing optical polarizers by integrating 2D $MoS_2$ films onto silicon photonic devices. Although other studies have reported higher *FOM* values using waveguides with larger cross-sections [61], thicker 2D materials [19], or enhanced mode overlap [14], it is important to highlight that our device achieves the highest *FOM* among all polarizers implemented based on the silicon photonic platform, which remains the most widely used and influential integrated photonics platform [63-65]. It is also worth noting that there remains substantial room to improve the *FOM* of $MoS_2$-Si hybrid waveguides by tailoring waveguide geometry to optimize mode overlap. Simulations show that a *FOM* value of ~19.2 can be achieved for a hybrid device with a cross-section of 400 nm × 160 nm for the bare silicon waveguide.

## VI. CONCLUSON

In summary, we integrate 2D $MoS_2$ films onto SOI nanowire waveguides to implement high-performance optical polarizers. High-quality monolayer $MoS_2$ films exhibiting highly anisotropic light absorption are synthesized via a LPCVD method, and then transferred onto SOI nanowire waveguides using a polymer-assisted transfer process. Detailed measurements are performed for our fabricated devices with various $MoS_2$ film coating lengths and silicon waveguide geometry. The results show that a maximum *PDL* of ~21 dB and a high *FOM* of ~4.2 are achieved. The hybrid waveguide polarizers also demonstrate a broad operation bandwidth of over ~100 nm and excellent power durability. These results verify the effectiveness of integrating 2D $MoS_2$ films onto silicon photonic devices for implementing high-performance optical polarizers.

REFERENCES

[1] Q. Bao, H. Zhang, B. Wang, Z. Ni, C. H. Y. X. Lim, Y. Wang, D. Y. Tang, and K. P. Loh, "Broadband graphene polarizer," *Nature Photonics,* vol. 5, no. 7, pp. 411-415, 2011/07/01, 2011.
[2] Y. Yan, G. Xie, M. P. J. Lavery, H. Huang, N. Ahmed, C. Bao, Y. Ren, Y. Cao, L. Li, Z. Zhao, A. F. Molisch, M. Tur, M. J. Padgett, and A. E. Willner, "High-capacity millimetre-wave communications with orbital angular momentum multiplexing," *Nature Communications,* vol. 5, no. 1, pp. 4876, 2014/09/16, 2014.
[3] D. Dai, L. Liu, S. Gao, D.-X. Xu, and S. He, "Polarization management for silicon photonic integrated circuits," *Laser & Photonics Reviews,* vol. 7, no. 3, pp. 303-328, 2013.
[4] K. Serkowski, D. S. Mathewson, and V. L. Ford, "Wavelength dependence of interstellar polarization and ratio of total to selective extinction," *The Astrophysical Journal,* vol. 196, pp. 261-290, February 01, 1975, 1975.
[5] T. J. Wang, Q. Y. He, J. Y. Gao, Y. Jiang, Z. H. Kang, H. Sun, L. S. Yu, X. F. Yuan, and J. Wu, "Efficient electrooptically Q-switched Er:Cr:YSGG laser oscillator-amplifier system with a Glan-Taylor prism polarizer," *Laser Physics,* vol. 16, no. 12, pp. 1605-1609, 2006/12/01, 2006.
[6] E. Saitoh, Y. Kawaguchi, K. Saitoh, and M. Koshiba, "TE/TM-Pass Polarizer Based on Lithium Niobate on Insulator Ridge Waveguide," *IEEE Photonics Journal,* vol. 5, no. 2, pp. 6600610-6600610, 2013.
[7] A. Rahnama, T. Dadalyan, K. Mahmoud Aghdami, T. Galstian, and P. R. Herman, "In-Fiber Switchable Polarization Filter Based on Liquid Crystal Filled Hollow-Filament Bragg Gratings," *Advanced Optical Materials,* vol. 9, no. 19, pp. 2100054, 2021.
[8] J. Wang, J.-Y. Yang, I. M. Fazal, N. Ahmed, Y. Yan, H. Huang, Y. Ren, Y. Yue, S. Dolinar, M. Tur, and A. E. Willner, "Terabit free-space data transmission employing orbital angular momentum multiplexing," *Nature Photonics,* vol. 6, no. 7, pp. 488-496, 2012/07/01, 2012.
[9] N. Bozinovic, Y. Yue, Y. Ren, M. Tur, P. Kristensen, H. Huang, A. E. Willner, and S. Ramachandran, "Terabit-Scale Orbital Angular Momentum Mode Division Multiplexing in Fibers," *Science,* vol. 340, no. 6140, pp. 1545-1548, 2013.
[10] D. Dai, Z. Wang, N. Julian, and J. E. Bowers, "Compact broadband polarizer based on shallowly-etched silicon-on-insulator ridge optical waveguides," *Optics Express,* vol. 18, no. 26, pp. 27404-27415, 2010/12/20, 2010.

[11] Y. Huang, S. Zhu, H. Zhang, T.-Y. Liow, and G.-Q. Lo, "CMOS compatible horizontal nanoplasmonic slot waveguides TE-pass polarizer on silicon-on-insulator platform," *Optics Express,* vol. 21, no. 10, pp. 12790-12796, 2013/05/20, 2013.

[12] J. Guo, Y. Liu, L. Lin, S. Li, J. Cai, J. Chen, W. Huang, Y. Lin, and J. Xu, "Chromatic Plasmonic Polarizer-Based Synapse for All-Optical Convolutional Neural Network," *Nano Letters,* vol. 23, no. 20, pp. 9651-9656, 2023/10/25, 2023.

[13] S. Wang, S. Wen, Z.-L. Deng, X. Li, and Y. Yang, "Metasurface-Based Solid Poincaré Sphere Polarizer," *Physical Review Letters,* vol. 130, no. 12, pp. 123801, 03/23/, 2023.

[14] H. Lin, Y. Song, Y. Huang, D. Kita, S. Deckoff-Jones, K. Wang, L. Li, J. Li, H. Zheng, Z. Luo, H. Wang, S. Novak, A. Yadav, C.-C. Huang, R.-J. Shiue, D. Englund, T. Gu, D. Hewak, K. Richardson, J. Kong, and J. Hu, "Chalcogenide glass-on-graphene photonics," *Nature Photonics,* vol. 11, no. 12, pp. 798-805, 2017/12/01, 2017.

[15] J. T. Kim, and H. Choi, "Polarization Control in Graphene-Based Polymer Waveguide Polarizer," *Laser & Photonics Reviews,* vol. 12, no. 10, pp. 1800142, 2018.

[16] J. T. Kim, and C.-G. Choi, "Graphene-based polymer waveguide polarizer," *Optics Express,* vol. 20, no. 4, pp. 3556-3562, 2012/02/13, 2012.

[17] J. Wu, Y. Yang, Y. Qu, X. Xu, Y. Liang, S. T. Chu, B. E. Little, R. Morandotti, B. Jia, and D. J. Moss, "Graphene Oxide Waveguide and Micro-Ring Resonator Polarizers," *Laser & Photonics Reviews,* vol. 13, no. 9, pp. 1900056, 2019.

[18] D. Jin, J. Wu, J. Hu, Y. Liu, Y. Zhang, Y. Yang, L. Jia, D. Huang, B. Jia, and D. J. Moss, "Silicon photonic waveguide and microring resonator polarizers incorporating 2D graphene oxide films," *Applied Physics Letters,* vol. 125, no. 5, 2024.

[19] W. H. Lim, Y. K. Yap, W. Y. Chong, C. H. Pua, N. M. Huang, R. M. De La Rue, and H. Ahmad, "Graphene oxide-based waveguide polariser: From thin film to quasi-bulk," *Optics Express,* vol. 22, no. 9, pp. 11090-11098, 2014/05/05, 2014.

[20] L. Zhuo, D. Li, W. Chen, Y. Zhang, W. Zhang, Z. Lin, H. Zheng, W. Zhu, Y. Zhong, J. Tang, G. Lu, W. Fang, J. Yu, and Z. Chen, "High performance multifunction-in-one optoelectronic device by integrating graphene/MoS2 heterostructures on side-polished fiber," *Nanophotonics,* vol. 11, no. 6, pp. 1137-1147, 2022.

[21] Y. Tan, R. He, C. Cheng, D. Wang, Y. Chen, and F. Chen, "Polarization-dependent optical absorption of MoS2 for refractive index sensing," *Scientific Reports,* vol. 4, no. 1, pp. 7523, 2014/12/17, 2014.

[22] S. Sathiyan, H. Ahmad, W. Y. Chong, S. H. Lee, and S. Sivabalan, "Evolution of the Polarizing Effect of MoS2," *IEEE Photonics Journal,* vol. 7, no. 6, pp. 1-10, 2015.

[23] Q. H. Wang, K. Kalantar-Zadeh, A. Kis, J. N. Coleman, and M. S. Strano, "Electronics and optoelectronics of two-dimensional transition metal dichalcogenides," *Nature Nanotechnology,* vol. 7, no. 11, pp. 699-712, 2012/11/01, 2012.

[24] F. Xia, H. Wang, D. Xiao, M. Dubey, and A. Ramasubramaniam, "Two-dimensional material nanophotonics," *Nature Photonics,* vol. 8, no. 12, pp. 899-907, 2014/12/01, 2014.

[25] S.-S. Chee, D. Seo, H. Kim, H. Jang, S. Lee, S. P. Moon, K. H. Lee, S. W. Kim, H. Choi, and M.-H. Ham, "Lowering the Schottky Barrier Height by Graphene/Ag Electrodes for High-Mobility MoS2 Field-Effect Transistors," *Advanced Materials,* vol. 31, no. 2, pp. 1804422, 2019.

[26] B. Radisavljevic, A. Radenovic, J. Brivio, V. Giacometti, and A. Kis, "Single-layer MoS2 transistors," *Nature Nanotechnology,* vol. 6, no. 3, pp. 147-150, 2011/03/01, 2011.

[27] J. Jiang, C. Ling, T. Xu, W. Wang, X. Niu, A. Zafar, Z. Yan, X. Wang, Y. You, L. Sun, J. Lu, J. Wang, and Z. Ni, "Defect Engineering for Modulating the Trap States in 2D Photoconductors," *Advanced Materials,* vol. 30, no. 40, pp. 1804332, 2018.

[28] O. Lopez-Sanchez, D. Lembke, M. Kayci, A. Radenovic, and A. Kis, "Ultrasensitive photodetectors based on monolayer MoS2," *Nature Nanotechnology,* vol. 8, no. 7, pp. 497-501, 2013/07/01, 2013.

[29] A. Hasani, M. Tekalgne, Q. V. Le, H. W. Jang, and S. Y. Kim, "Two-dimensional materials as catalysts for solar fuels: hydrogen evolution reaction and CO2 reduction," *Journal of Materials Chemistry A,* vol. 7, no. 2, pp. 430-454, 2019.

[30] M. Asadi, K. Kim, C. Liu, A. V. Addepalli, P. Abbasi, P. Yasaei, P. Phillips, A. Behranginia, J. M. Cerrato, R. Haasch, P. Zapol, B. Kumar, R. F. Klie, J. Abiade, L. A. Curtiss, and A. Salehi-Khojin, "Nanostructured transition metal dichalcogenide electrocatalysts for $CO_2$ reduction in ionic liquid," *Science,* vol. 353, no. 6298, pp. 467-470, 2016.

[31] S. Presolski, and M. Pumera, "Covalent functionalization of MoS2," *Materials Today,* vol. 19, no. 3, pp. 140-145, 2016/04/01/, 2016.

[32] D.-M. Tang, D. G. Kvashnin, S. Najmaei, Y. Bando, K. Kimoto, P. Koskinen, P. M. Ajayan, B. I. Yakobson, P. B. Sorokin, J. Lou, and D. Golberg, "Nanomechanical cleavage of molybdenum disulphide atomic layers," *Nature Communications,* vol. 5, no. 1, pp. 3631, 2014/04/03, 2014.

[33] K. F. Mak, C. Lee, J. Hone, J. Shan, and T. F. Heinz, "Atomically Thin MoS2: A New Direct-Gap Semiconductor," *Physical Review Letters,* vol. 105, no. 13, pp. 136805, 09/24/, 2010.

[34] I. H. Abidi, S. P. Giridhar, J. O. Tollerud, J. Limb, M. Waqar, A. Mazumder, E. L. Mayes, B. J. Murdoch, C. Xu, A. Bhoriya, A. Ranjan, T. Ahmed, Y. Li, J. A. Davis, C. L. Bentley, P. R. Russo, E. D. Gaspera, and S. Walia, "Oxygen Driven Defect Engineering of Monolayer MoS2 for Tunable Electronic, Optoelectronic, and Electrochemical Devices," *Advanced Functional Materials,* vol. 34, no. 37, pp. 2402402, 2024.

[35] I. H. Abidi, A. Bhoriya, P. Vashishtha, S. P. Giridhar, E. L. H. Mayes, M. Sehrawat, A. K. Verma, V. Aggarwal, T. Gupta, H. K. Singh, T. Ahmed, N. Dilawar Sharma, and S. Walia, "Oxidation-induced modulation of photoresponsivity in monolayer MoS2 with sulfur vacancies," *Nanoscale,* vol. 16, no. 42, pp. 19834-19843, 2024.

[36] K. M. Islam, R. Synowicki, T. Ismael, I. Oguntoye, N. Grinalds, and M. D. Escarra, "In-Plane and Out-of-Plane Optical Properties of Monolayer, Few-Layer, and Thin-Film MoS2 from 190 to 1700 nm and Their Application in Photonic Device Design," *Advanced Photonics Research,* vol. 2, no. 5, pp. 2000180, 2021.

[37] G. A. Ermolaev, D. V. Grudinin, Y. V. Stebunov, K. V. Voronin, V. G. Kravets, J. Duan, A. B. Mazitov, G. I. Tselikov, A. Bylinkin, D. I. Yakubovsky, S. M. Novikov, D. G. Baranov, A. Y. Nikitin, I. A. Kruglov, T. Shegai, P. Alonso-González, A. N. Grigorenko, A. V. Arsenin, K. S. Novoselov, and V. S. Volkov, "Giant optical anisotropy in transition metal dichalcogenides for next-generation photonics," *Nature Communications,* vol. 12, no. 1, pp. 854, 2021/02/08, 2021.

[38] Q. Feng, H. Cong, B. Zhang, W. Wei, Y. Liang, S. Fang, T. Wang, and J. Zhang, "Enhanced optical Kerr nonlinearity of graphene/Si hybrid waveguide," *Applied Physics Letters,* vol. 114, no. 7, 2019.

[39] H. Cai, Y. Cheng, H. Zhang, Q. Huang, J. Xia, R. Barille, and Y. Wang, "Enhanced linear absorption coefficient of in-plane monolayer graphene on a silicon microring resonator," *Optics Express,* vol. 24, no. 21, pp. 24105-24116, 2016/10/17, 2016.

[40] C. Martella, C. Mennucci, E. Cinquanta, A. Lamperti, E. Cappelluti, F. Buatier de Mongeot, and A. Molle, "Anisotropic MoS2 Nanosheets Grown on Self-Organized Nanopatterned Substrates," *Advanced Materials,* vol. 29, no. 19, pp. 1605785, 2017.

[41] G.-H. Nam, Q. He, X. Wang, Y. Yu, J. Chen, K. Zhang, Z. Yang, D. Hu, Z. Lai, B. Li, Q. Xiong, Q. Zhang, L. Gu, and H. Zhang, "In-Plane Anisotropic Properties of 1T′-MoS2 Layers," *Advanced Materials,* vol. 31, no. 21, pp. 1807764, 2019.

[42] Y. Zhang, J. Wu, L. Jia, D. Jin, B. Jia, X. Hu, D. Moss, and Q. Gong, "Advanced optical polarizers based on 2D materials," *npj Nanophotonics,* vol. 1, no. 1, pp. 28, 2024/07/17, 2024.

[43] L. Jia, J. Wu, Y. Zhang, Y. Qu, B. Jia, Z. Chen, and D. J. Moss, "Fabrication Technologies for the On-Chip Integration of 2D Materials," *Small Methods,* vol. 6, no. 3, pp. 2101435, 2022.

[44] A. Gurarslan, Y. Yu, L. Su, Y. Yu, F. Suarez, S. Yao, Y. Zhu, M. Ozturk, Y. Zhang, and L. Cao, "Surface-Energy-Assisted Perfect Transfer of Centimeter-Scale Monolayer and Few-Layer MoS2 Films onto Arbitrary Substrates," *ACS Nano,* vol. 8, no. 11, pp. 11522-11528, 2014/11/25, 2014.

[45] W. M. Parkin, A. Balan, L. Liang, P. M. Das, M. Lamparski, C. H. Naylor, J. A. Rodríguez-Manzo, A. T. C. Johnson, V. Meunier, and M. Drndić, "Raman Shifts in Electron-Irradiated Monolayer MoS2," *ACS Nano,* vol. 10, no. 4, pp. 4134-4142, 2016/04/26, 2016.

[46] J. Zhang, H. Yu, W. Chen, X. Tian, D. Liu, M. Cheng, G. Xie, W. Yang, R. Yang, X. Bai, D. Shi, and G. Zhang, "Scalable Growth of High-Quality Polycrystalline MoS2 Monolayers on SiO2 with Tunable Grain Sizes," *ACS Nano,* vol. 8, no. 6, pp. 6024-6030, 2014/06/24, 2014.

[47] L. Britnell, R. M. Ribeiro, A. Eckmann, R. Jalil, B. D. Belle, A. Mishchenko, Y.-J. Kim, R. V. Gorbachev, T. Georgiou, S. V. Morozov, A. N. Grigorenko, A. K. Geim, C. Casiraghi, A. H. C. Neto, and K. S.

Novoselov, "Strong Light-Matter Interactions in Heterostructures of Atomically Thin Films," *Science,* vol. 340, no. 6138, pp. 1311-1314, 2013.

[48] D. Dumcenco, D. Ovchinnikov, K. Marinov, P. Lazić, M. Gibertini, N. Marzari, O. L. Sanchez, Y.-C. Kung, D. Krasnozhon, M.-W. Chen, S. Bertolazzi, P. Gillet, A. Fontcuberta i Morral, A. Radenovic, and A. Kis, "Large-Area Epitaxial Monolayer MoS2," *ACS Nano,* vol. 9, no. 4, pp. 4611-4620, 2015/04/28, 2015.

[49] V. Forsberg, R. Zhang, J. Bäckström, C. Dahlström, B. Andres, M. Norgren, M. Andersson, M. Hummelgård, and H. Olin, "Exfoliated MoS2 in Water without Additives," *PLOS ONE,* vol. 11, no. 4, pp. e0154522, 2016.

[50] X. Liu, T. Wang, G. Hu, C. Xu, Y. Xiong, and Y. Wang, "Controllable synthesis of self-assembled MoS2 hollow spheres for photocatalytic application," *Journal of Materials Science: Materials in Electronics,* vol. 29, no. 1, pp. 753-761, 2018/01/01, 2018.

[51] H. Liu, D. Su, R. Zhou, B. Sun, G. Wang, and S. Z. Qiao, "Highly Ordered Mesoporous MoS2 with Expanded Spacing of the (002) Crystal Plane for Ultrafast Lithium Ion Storage," *Advanced Energy Materials,* vol. 2, no. 8, pp. 970-975, 2012.

[52] C. P. Veeramalai, F. Li, Y. Liu, Z. Xu, T. Guo, and T. W. Kim, "Enhanced field emission properties of molybdenum disulphide few layer nanosheets synthesized by hydrothermal method," *Applied Surface Science,* vol. 389, pp. 1017-1022, 2016/12/15/, 2016.

[53] H. Nan, Z. Wang, W. Wang, Z. Liang, Y. Lu, Q. Chen, D. He, P. Tan, F. Miao, X. Wang, J. Wang, and Z. Ni, "Strong Photoluminescence Enhancement of MoS2 through Defect Engineering and Oxygen Bonding," *ACS Nano,* vol. 8, no. 6, pp. 5738-5745, 2014/06/24, 2014.

[54] S.-S. Chee, W.-J. Lee, Y.-R. Jo, M. K. Cho, D. Chun, H. Baik, B.-J. Kim, M.-H. Yoon, K. Lee, and M.-H. Ham, "Atomic Vacancy Control and Elemental Substitution in a Monolayer Molybdenum Disulfide for High Performance Optoelectronic Device Arrays," *Advanced Functional Materials,* vol. 30, no. 11, pp. 1908147, 2020.

[55] N. Jiménez-Arévalo, J. H. Al Shuhaib, R. B. Pacheco, D. Marchiani, M. M. Saad Abdelnabi, R. Frisenda, M. Sbroscia, M. G. Betti, C. Mariani, Y. Manzanares-Negro, C. G. Navarro, A. J. Martínez-Galera, J. R. Ares, I. J. Ferrer, and F. Leardini, "MoS2 Photoelectrodes for Hydrogen Production: Tuning the S-Vacancy Content in Highly Homogeneous Ultrathin Nanocrystals," *ACS Applied Materials & Interfaces,* vol. 15, no. 28, pp. 33514-33524, 2023/07/19, 2023.

[56] D. Dai, J. Bauters, and J. E. Bowers, "Passive technologies for future large-scale photonic integrated circuits on silicon: polarization handling, light non-reciprocity and loss reduction," *Light: Science & Applications,* vol. 1, no. 3, pp. e1-e1, 2012/03/01, 2012.

[57] J. Wu, Y. Zhang, J. Hu, Y. Yang, D. Jin, W. Liu, D. Huang, B. Jia, and D. J. Moss, "2D graphene oxide films expand functionality of photonic chips," *Advanced Materials*, pp. 2403659, 2024.

[58] Y. Zhang, L. Tao, D. Yi, J.-b. Xu, and H. K. Tsang, "Enhanced thermo-optic nonlinearities in a MoS2-on-silicon microring resonator," *Applied Physics Express,* vol. 13, no. 2, pp. 022004, 2020/01/08, 2020.

[59] H. Li, Y. Anugrah, S. J. Koester, and M. Li, "Optical absorption in graphene integrated on silicon waveguides," *Applied Physics Letters,* vol. 101, no. 11, 2012.

[60] Y. Zhang, J. Wu, Y. Yang, Y. Qu, L. Jia, T. Moein, B. Jia, and D. J. Moss, "Enhanced Kerr Nonlinearity and Nonlinear Figure of Merit in Silicon Nanowires Integrated with 2D Graphene Oxide Films," *ACS Applied Materials & Interfaces,* vol. 12, no. 29, pp. 33094-33103, 2020/07/22, 2020.

[61] C. Pei, L. Yang, G. Wang, Y. Wang, X. Jiang, Y. Hao, Y. Li, and J. Yang, "Broadband Graphene/Glass Hybrid Waveguide Polarizer," *IEEE Photonics Technology Letters,* vol. 27, no. 9, pp. 927-930, 2015.

[62] J. Hu, J. Wu, D. Jin, W. Liu, Y. Zhang, Y. Yang, L. Jia, Y. Wang, D. Huang, B. Jia, and D. J. Moss, "Integrated photonic polarizers with 2D reduced graphene oxide," *Opto-Electronic Science*, pp. 240032, 2025/02/26, 2025.

[63] A. Rickman, "The commercialization of silicon photonics," *Nature Photonics,* vol. 8, no. 8, pp. 579-582, 2014/08/01, 2014.

[64] J. Leuthold, C. Koos, and W. Freude, "Nonlinear silicon photonics," *Nature Photonics,* vol. 4, no. 8, pp. 535-544, 2010/08/01, 2010.

[65] D. J. Moss, R. Morandotti, A. L. Gaeta, and M. Lipson, "New CMOS-compatible platforms based on silicon nitride and Hydex for nonlinear optics," *Nature Photonics,* vol. 7, no. 8, pp. 597-607, 2013/08/01, 2013.

**Junkai Hu** (Member, IEEE) received his bachelor's degree from Central South University, China in June 2020. He is currently a Ph.D. candidate in Central South University and visiting Ph.D. student at the Optical Sciences Centre in Swinburne University of Technology under the supervision of Prof. David Moss and Dr. Jiayang Wu. His current research focuses on functional integrated photonic devices incorporating 2D materials.

**Jiayang Wu** (Senior Member, IEEE) received the B.Eng. degree in September 2010 from Xidian University, Xi'an, China, and the Ph.D. degree in December 2015 from Shanghai Jiao Tong University, Shanghai, China. After that, he joined the Swinburne University of Technology and became a Postdoctoral Research Fellow in 2016. He is currently a Senior Research Fellow and Senior Lecturer at the Optical Sciences Centre, Swinburne University of Technology. His current research fields include integrated photonics, nonlinear optics, and 2D materials. As of May 1st of 2025, he has over 100 publications in SCI journals, highlighted by *Nature*, *Nature Reviews Chemistry*, *Advanced Materials*, *Advances in Optics and Photonics*, *Nature Communications*, *Applied Physics Reviews*, *Light: Advanced Manufacturing*, *Nano Letters*, *Small*, and *Laser & Photonics Review*. He is also an inventor on 15 filed technological patents. In 2021, Dr. Wu was awarded the Australian National Research Award as the top researcher in the field of Optics & Photonics (only 1 person). From 2021 to 2024, He was consecutively named in the world's top 2% of scientists list (*Stanford University & Elsevier science-wide author databases of standardized citation indicators*).

**Irfan H. Abidi** received the Ph.D. degree in Chemical and Biomolecular Engineering (Nanotechnology) from the Hong Kong University of Science and Technology (HKUST), Hong Kong, in 2018. From 2018 to 2021, he was a Research Fellow with the Centre for Advanced 2D Materials, National University of Singapore, a world-leading graphene research center. He is currently a Research Fellow with the School of Engineering, RMIT University, Australia. His research focuses on engineering atomic structures in 2D materials from single crystalline to amorphous phases, using chemical vapor deposition (CVD) and laser-CVD techniques for applications in optoelectronics, quantum technologies, energy storage, and electrochemical devices. His pioneering contributions include the "gettering" CVD growth of high-quality monolayer graphene and h-BN, and the first-ever synthesis of monolayer amorphous carbon. He has co-authored over 40 peer-reviewed publications, including articles in *Nature*, *Advanced Functional Materials*, and *ACS Nano*, is a co-inventor on four U.S. patents, and serves as editor of a scientific book published by Elsevier. His work bridges fundamental science and applied research, addressing key industrial challenges through strong collaborations with national and international academic and industry partners.

**Di Jin** (Member, IEEE) received her bachelor's degree from Central South University, China in June 2017. She is currently a Ph.D. candidate in Central South University and visiting Ph.D. student at the Optical Sciences Centre in Swinburne University of Technology under the supervision of Prof. David Moss and Dr. Jiayang Wu. Her current research focuses on 2D materials, nonlinear optics, and integrated photonics.

**Yuning Zhang** (Member, IEEE) received her bachelor's degree from Tianjin University of Science and Technology, China in June 2013 and her master's degree from Beihang University, China and her Ph.D. degree from Swinburne University of Technology, Australia in 2013, 2017, 2022, respectively. She is currently a PhD candidate at the Optical Sciences Centre in Swinburne University of Technology under the supervision of Prof. David Moss and Dr. Jiayang Wu. Her current research focuses on functional integrated photonic devices incorporating 2D materials.

**Jianfeng Mao** received the B.Eng. degree in Materials Science and Engineering from Northwestern Polytechnical University, Xi'an, China, in 2015, and the M.Eng. degree in Materials Science and Engineering from Shanghai Jiao Tong University, Shanghai, China, in 2018. He earned his Ph.D. in Applied Physics from The Hong Kong Polytechnic University, Hong Kong, in 2024. Following his doctoral studies, he worked as a Postdoctoral Research Fellow in the Department of Applied Physics at Hong Kong Polytechnic University from April to October 2024. He is currently a Postdoctoral Research Fellow at the School of Engineering, RMIT University, Australia. Dr. Mao's research focuses on two-dimensional materials and their strain-modulated optoelectronic and ferroelectric properties. As of 2025, Dr. Mao has authored over 11 peer-reviewed journal papers in leading journals such as *Nature Communications*, *Advanced Materials*, *ACS Nano*, *Nano Research*, and *ACS Applied Materials & Interfaces*.

**Anchal Pandey** is a Ph.D. candidate in the joint AcSIR-RMIT program and currently works in the School of Engineering at RMIT University, Australia. She received her master's degree in chemistry from the University of Allahabad, India. Her doctoral research focuses on atomic-scale friction in two-dimensional heterostructures, examining the role of interfacial lattice mismatch using tribological and surface characterization techniques. Her work aims to advance the synthesis of ultra-low friction lubricants and deepen understanding of interfacial mechanics in 2D systems. She has published in peer-reviewed journals and is actively involved in interdisciplinary research and scientific outreach.

**Yijun Wang** received his bachelor's degree in road and railway engineering from Changsha Railway University, China and his master's degree in civil engineering and his Ph.D. degree in control science and engineering from Central South University, China in 1985, 2003 and 2009, respectively. Since 2003, he has been a Professor with the School of Information Science and Engineering, Central South University. He is the author of more than 50 articles. His current research focuses on railway informatization, transportation information engineering and control, and quantum information technology.

**Sumeet Walia** is a Professor and Director of RMITs Centre for Opto-electronic materials and sensors (COMAS). His research focuses on discovering and manipulating fundamental properties of materials for applications across energy, nano/optoelectronics, sensors and healthcare. He has a demonstrated track of creating bespoke, needs-based multidisciplinary teams to solve critical technological bottlenecks. He has co-authored over 160 peer reviewed publications including authoritative reviews in prestigious journals, a named inventor on twelve patents and editor of two books for the CRC Press. Sumeet partners with several cross-sector industry to translate fundamental discoveries into the real world. He is also an avid contributor in enhancing Equity, Diversity, Inclusion and Access in STEM. He chaired the national equity, diversity and inclusion committee of Science and Technology Australia and has been part of advisory boards of the Victorian government and the Australian Academy of Technology and Engineering. His scientific and leadership contributions have been recognised through several national and international awards including the Eureka Prize for Science leadership and the MIT Technology Review's Top 10 Innovators in APAC.

**David J. Moss** (Life Fellow, IEEE) is Director of the Optical Sciences Centre at Swinburne University of Technology in Melbourne, Australia, and Deputy Director of the ARC Centre of Excellence for optical microcombs for breakthrough science. He was with RMIT University in Melbourne from 2014 to 2016, the University of Sydney from 2004 to 2014, and JDSUniphase in Ottawa Canada from 1998 to 2003. From 1994 to 1998 he was with the Optical Fiber Technology Centre at Sydney University, from 1992 ro 1994 with Hitachi Central Research Laboratories in Tokyo, Japan, and from 1988 to 1992 at the National Research Council of Canada in Ottawa. He received his Ph.D. from the University of Toronto and B.Sc. from the University of Waterloo. He won the 2011 Australian Museum Eureka Science Prize and Google Australia Prize for Innovation in Computer Science. He is a Life Fellow of the IEEE Photonics Society, Life Fellow of Optica (formerly the OSA), and Life Fellow of the SPIE. He is a Fellow of the Royal Society of Canada (RSC) and the Australian Academy of Technological Sciences and Engineering (ATSE). In 2024 he was awarded an honorary doctorate (honoris causa) from the Danish Technical University (DTU) and King Frederik X of Denmark for the invention and pioneering of optical microcombs. His research interests include optical microcombs, integrated nonlinear optics, quantum optics, microwave photonics, ONNs, optical networks and transmission, 2D materials for nonlinear optics, optical signal processing, nanophotonics, and biomedical photonics for cancer diagnosis and therapy.